# Thin films of metallic carbon nanotubes and their optical spectra

Ralph Krupke[1], Stefan Linden[1,2], Michael Rapp[3], Frank Hennrich[1]

[1]*Forschungszentrum Karlsruhe, Institut für Nanotechnologie, D-76021 Karlsruhe*

[2]*Institut für angewandte Physik, Universität Karlsruhe, D-76128 Karlsruhe*

[3]*Forschungszentrum Karlsruhe, Institut für Instrumentelle Analytik, D-76021 Karlsruhe*

We show, that separating metallic from semiconducting carbon nanotubes by dielectrophoresis is developing towards a bulk separation method, which allows for the first time to produce thin films of only metallic single-walled carbon nanotubes and to measure their optical absorption spectra. The data proofs that the selectivity of the separation scheme is independent from the nanotube diameter.

The unique electronic and structural properties of metallic and semiconducting single-walled carbon nanotubes (SWNTs) have lead to the vision of carbon-nanotube based transistors or interconnects for future nanoscale electronics [1]. Such potential applications, however, are challenged by the inability to selectively produce metallic or semiconducting SWNTs, instead of a mixture. Over the last two years, progress has been made by exploring various concepts for separation of metallic from semiconducting SWNTs [2]. So far, chemical routes, which could easily be scaled up, do not provide full spatial separation, but rather selective enrichment. On the other hand, dielectrophoresis, a physical process, has been shown to separate metallic from semiconducting tubes with high selectivity but low yield [3]. Here we show that dielectrophoresis is developing towards a bulk separation method, allowing for the first time to produce thin films of only metallic SWNTs and to measure the corresponding optical spectra.

Dielectrophoresis (DEP), which describes the motion of a polarizable object due to an external inhomogeneous electrical field, has become an important sorting mechanism in life sciences and allows to sort cells and viruses by their dielectric properties [4,5]. DEP forces are inherently weak for small particles due to scaling with the particle volume. For SWNTs, with typical dimensions of 1 nm in diameter times 1 μm in length, the DEP forces are about three orders of magnitude smaller as compared to a 100 nm particle of similar polarizability. To observe nonetheless DEP sorting of SWNTs, the electrical field strength and gradient must be respectively large. Moreover electrophoretic motion, due to particle charges, must be inhibited. We demonstrated earlier by Raman spectroscopy, that metallic tubes can indeed be separated from semiconducting SWNTs by DEP deposition from a SWNT suspension using radio-frequency electrical fields [3]. The selectivity of the process depends on the field strength and frequency, on the surfactant concentration or solution conductivity, and on the content of SWNTs that are not individually dispersed but aggregated in bundles [6]. Improving the DEP nanotube separation process allowed us now to sort out metallic tubes with a three order of magnitude higher yield and to measure for the first time optical absorption spectra of a thin film of exclusively metallic carbon nanotubes. For the purpose of the experiment we have deposited laser-ablated SWNTs from surfactant-stabilized aqueous suspension onto a transparent glass substrate and measured the optical absorbance in transmission with a Fourier-transform infrared spectrometer combined with an infrared microscope and using unpolarised light. The SWNT suspensions were prepared by sonicating the raw material in $D_2O$ with 0.1 or 1 wt % sodium cholate and by subsequent centrifugation at 154kg for 2 h. The carefully decanted supernatant contained individually dispersed SWNTs with a concentration of about 10 ng / µl. The array of interdigitated electrodes was prepared by optical lithography on a quartz glass substrate with the gap between the fingers of 1.8 µm. DEP deposition was achieved by driving the array by a radio-frequency generator at a voltage of $V_{pp}$ = 20 V for 5 minutes during exposure to a drop of 10 µl SWNT suspension. Subsequently the sample was rinsed with $H_2O$ and dried in a stream of nitrogen gas.

Figure 1 shows the morphology of a DEP deposited film and an illustration of the experimental setup. The individually dispersed SWNTs agglomerate during the DEP deposition to bundles on the surface and form a film with the preferential alignment being perpendicular to the electrode fingers. The optical absorption spectra of the films formed at various electric field frequencies and surfactant concentrations are presented in Figure 2. The spectral window has been chosen such, that the absorption measurements cover the S2 absorption band of semiconducting SWNTs and the M1 absorption band of metallic SWNTs. The prime spectroscopic result is the observation, that the S2 absorption band is absent in thin nanotube films, prepared by DEP at high electric-field frequency and at high surfactant concentration. At the same time the M1 absorption band remains fully developed. The data is evidence that only metallic tubes were deposited onto the glass surface and thereby separated form the semiconducting tubes - which remain in the suspension. Since the absorption band width is determined by the nanotube diameter distribution, the data shows furthermore that separation is achieved for nanotubes with various diameters. Otherwise the spectral shape of the S2 band would have changed. That the technique is apparently independent from the diameter distribution of the starting material is in agreement with our former successful separation of HiPco tubes. The data also shows that no separation is achieved at surfactant concentrations below the critical micelle concentration or at low electric-field frequency. We believe that this effect is related to the residual bundle content in tube suspensions at low surfactant concentration and to a surfactant induced surface-conductance phenomena, as we have pointed out in an earlier publication [6]. The details are currently under investigation. We would like to add that the absorption spectra were not sensitive to thermal annealing in vacuum up to 600° C.

We have estimated the thickness of the DEP deposited films to be on the order of 10 nm, by comparing the optical density (OD) of the films with the OD of a freestanding nanotube bucky-paper of known thickness [7]. Supposing a similar weight density for the DEP-films and for the bucky paper of 1 g / cm$^3$, we derive the amount of DEP deposited material to be of the order of 1 μg / cm$^2$. Note that this DEP-film thickness equals the amount of tubes in the volume of a 1 mm thick suspension layer above the electrode array and thereby suggest a rather efficient sorting if we take into account that the DEP forces exponentially decay above the electrode array [8].

We conclude that dielectrophoretic separation of metallic from semiconducting tubes on the basis of their different dielectric properties, is developing towards a bulk separation method, without sacrificing the intrinsic high selectivity of the process – an advancement, which is promising for the development of nanotube-based electronic device applications.

**Acknowledgements.** R.K.acknowledges funding by the Initiative and Networking Fund of the Helmholtz-Gemeinschaft Deutscher Forschungszentren (HGF).

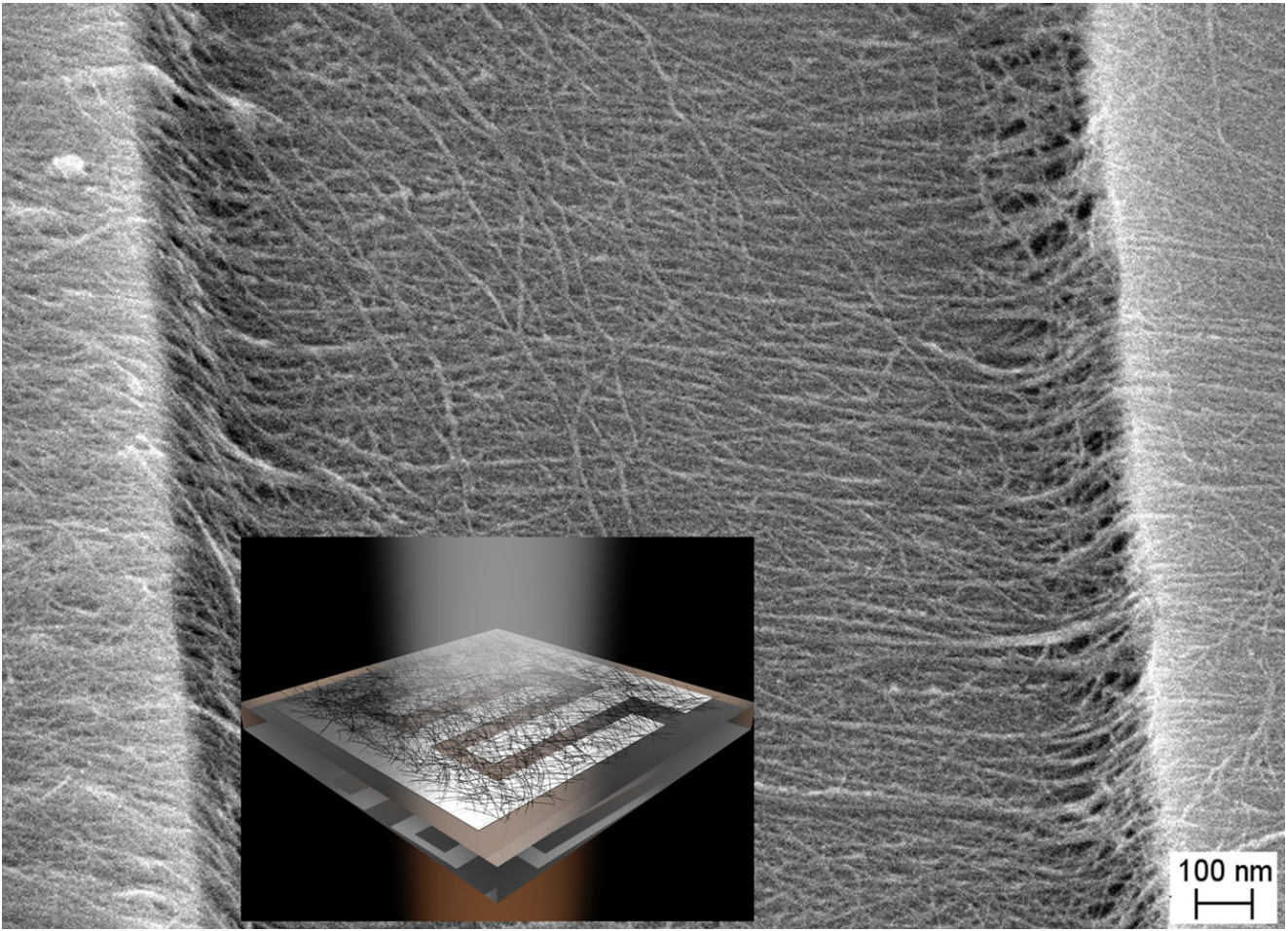

**Fig. 1.** Electron microscopy image of a thin film formed on a transparent glass substrate by dielectrophoresis from a suspension of individually dispersed single-walled carbon nanotubes (SWNT). The electric field strength, generated during deposition by the 1.8 µm gapped interdigitated electrodes, is of the order of $10^7$ V / m. Local absorption spectroscopy is performed in transmission through a 200 x 200 (µm)$^2$ area with unpolarised light (inset).

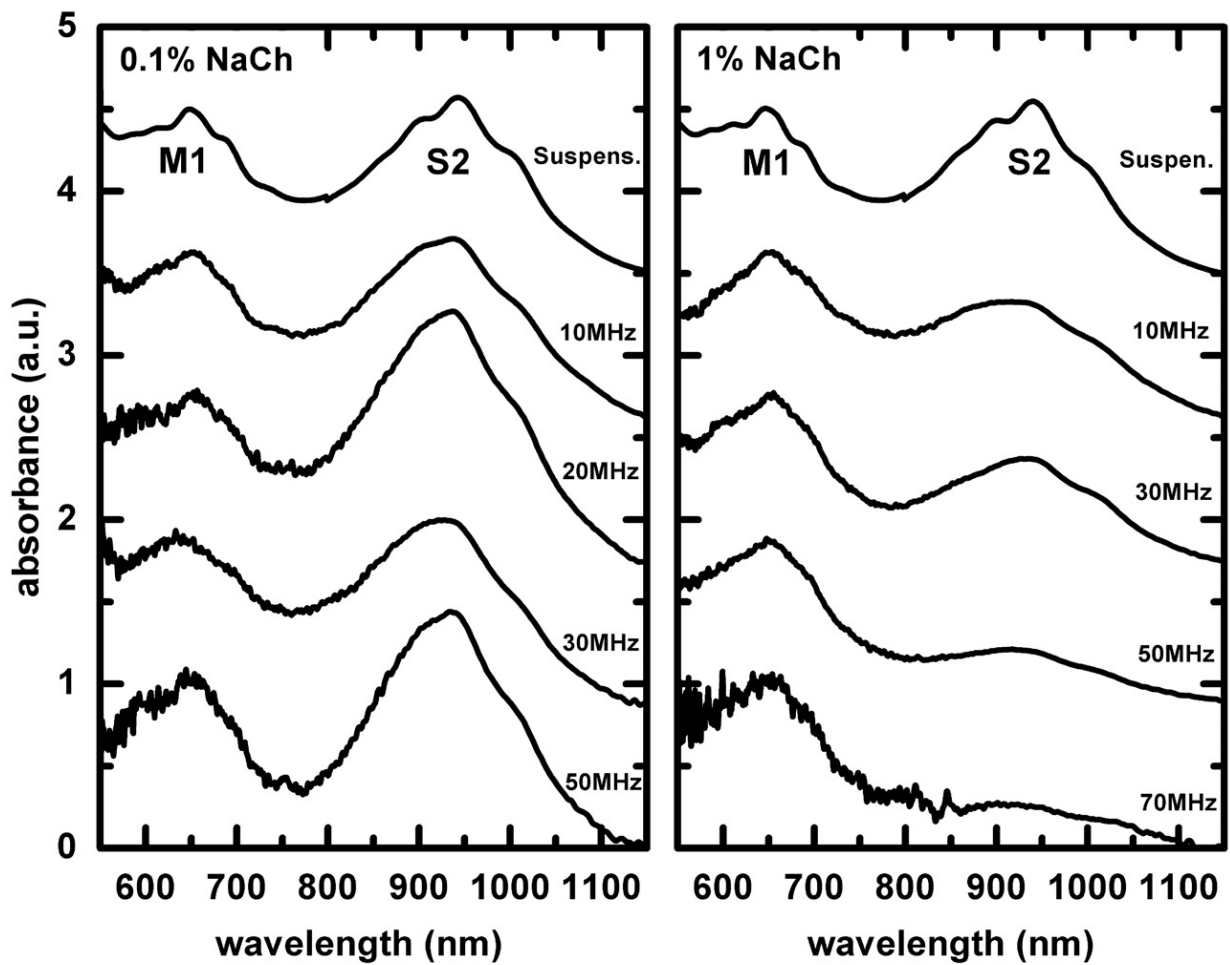

**Fig. 2**. Optical absorption spectra of films, formed by dielectrophoresis from SWNT suspensions with 0.1 % and 1 % of sodium cholate (NaCh) surfactant. Electric field frequencies are ranging from 10 MHz to 70 MHz. Also given are the spectra of the starting suspensions. Absorption bands of semiconducting SWNTs (S2) and of metallic SWNTs (M1) are indicated. The data has been normalized to the maximum of the M1 band and has been shifted vertically for clarity. The SWNT film formed at high electric-field frequency from a 1 % NaCh suspension is composed of primarily metallic SWNTs.